\documentclass[aps,prl,reprint,twocolumn,superscriptaddress,tightenlines,amsmath, amssymb]{revtex4-1}
\usepackage{mathptmx}
\usepackage[dvips]{color}
\usepackage[dvips]{graphicx}

\makeatletter

\@ifundefined{textcolor}{}
{%
 \definecolor{BLACK}{cmyk}{0,0,0,1}
 \definecolor{WHITE}{gray}{1}
 \definecolor{RED}{cmyk}{0,1,1,0}
 \definecolor{GREEN}{cmyk}{1,0,1,0.5}
 \definecolor{BLUE}{cmyk}{1,1,0,0}
 \definecolor{CYAN}{cmyk}{1,0,0,0}
 \definecolor{MAGENTA}{cmyk}{0,1,0,0}
 \definecolor{YELLOW}{cmyk}{0,0,1,0}
  \definecolor{ORANGE}{cmyk}{0,0.75,0.25,0}
}

\makeatother

\begin{document}

\title{The Marangoni flow of soluble amphiphiles}

\author{Matthieu Roch\'{e}}
\altaffiliation[Now at: ]{Laboratoire de Physique des Solides, Universit\'{e} Paris Sud - CNRS UMR 8502, B\^{a}timent 510, 91405 Orsay, France}
\email[E-mail address: ]{matthieu.roche@u-psud.fr}
\affiliation{Department of Mechanical and Aerospace Engineering, Princeton University,
Princeton, New Jersey 08544, USA}

\author{Zhenzhen Li}
\altaffiliation[Now at: ]{MMN, CNRS, ESPCI Paris-Tech, 10 rue Vauquelin, 75005 Paris, France}
\affiliation{Department of Mechanical and Aerospace Engineering, Princeton University,
Princeton, New Jersey 08544, USA}

\author{Ian M. Griffiths}
\affiliation{Mathematical Institute, University of Oxford, Oxford, OX2 6GG, England}

\author{S\'{e}bastien Le Roux}
\affiliation{Institut de Physique de Rennes, CNRS UMR 6251, Universit\'{e} Rennes 1, 35042 Rennes, France}
\author{Isabelle Cantat}
\affiliation{Institut de Physique de Rennes, CNRS UMR 6251, Universit\'{e} Rennes 1, 35042 Rennes, France}
\author{Arnaud Saint-Jalmes}
\affiliation{Institut de Physique de Rennes, CNRS UMR 6251, Universit\'{e} Rennes 1, 35042 Rennes, France}
\author{Howard A. Stone}
\affiliation{Department of Mechanical and Aerospace Engineering, Princeton University,
Princeton, New Jersey 08544, USA}

\begin{abstract}
Surfactant distribution heterogeneities at a fluid/fluid interface trigger the Marangoni effect, \textit{i.e.} a bulk flow due to a surface tension gradient. The influence of surfactant solubility in the bulk on these flows remains incompletely characterized. Here we study Marangoni flows sustained by injection of hydrosoluble surfactants at the air/water interface. We show that the flow extent increases with a decrease of the critical micelle concentration, \textit{i.e.} the concentration at which these surfactants self-assemble in water. We document the universality of the surface velocity field and predict scaling laws based on hydrodynamics and surfactant physicochemistry that capture the flow features.
\end{abstract}

\maketitle
The release of a drop of water mixed with dishwashing liquid on the surface of pure water covered with pepper grains demonstrates the Marangoni effect \cite{Levich1969}: after the drop touches the water surface, pepper grains are transported rapidly to the edge of the bowl (see movie M1 in Supplementary Materials). The flow results from the difference in surface tension between water at the point of release and clean water far away. The occurrence of the Marangoni effect plays an important role in many natural and industrial processes such as pulmonary surfactant replacement therapy \cite{Jobe1993,Grotberg1994}, motion and defence of living organisms \cite{Harshey2003,Eisner2005,Bush2006}, the stability of emulsions and foams \cite{Ivanov1997,Cantat2013} and many others \cite{Quere1999,Matar2001,Barnes2008,Herzig2011}. In these settings, surfactants generally have a finite solubility in one of the phases, but the effect of interface-bulk mass exchange on Marangoni flows is still not understood despite its consequences on flow (see Movie M2 in Supplementary Materials). Most studies \cite{Halpern1992,Jensen1993,Lee2007,Lee2009} have focused on the deposition of droplets of surfactant solutions on thin water films, but the transient nature of the induced flow and the small size of the film prevented the validation of proposed theoretical descriptions. Here, we investigate axisymmetric Marangoni flows induced by hydrosoluble surfactants on centimeter-thick water layers. We show how the flow extent and the associated velocity field depend on the surfactant chemical structure, hence on amphiphile thermodynamic properties such as the critical micelle concentration (cmc). We propose scaling laws based on hydrodynamics and physicochemistry for the flow radius and its velocity that are in excellent agreement with the experiments.

\begin{figure*}[!ht]
	\includegraphics[scale=1]{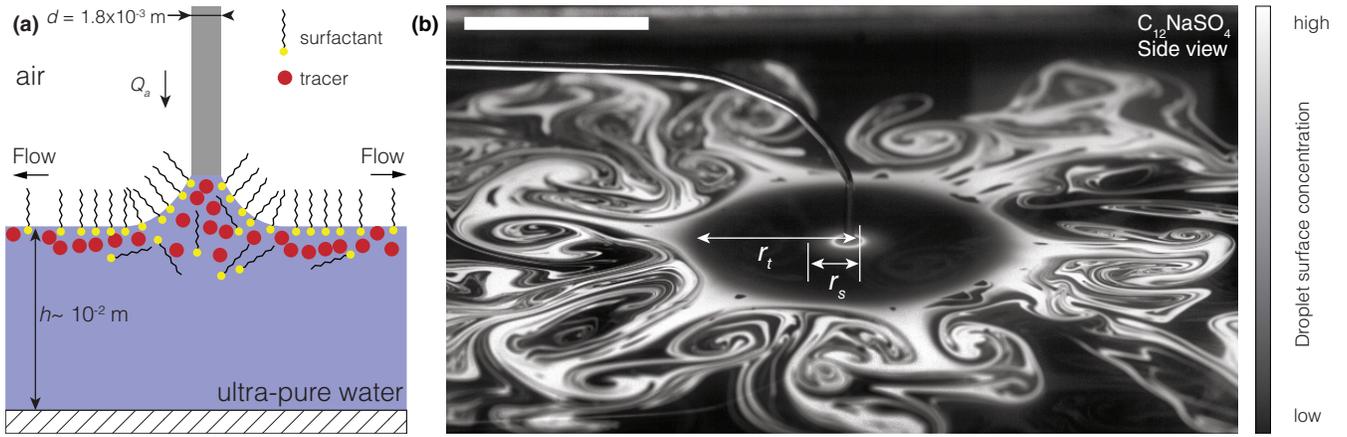}
	\caption{\textbf{Experimental observation of the continuous surfactant-induced Marangoni flow.} (a) Schematic of the experimental set-up in the region surrounding the point of injection. (b) Side view of a typical experiment. In this view, regions of high coverage in light-scattering tracers are white. Surfactant molar flow rate $Q_{a}= 0.52\times10^{-6}$ mol.s$^{-1}$, scale bar: 3$\times10^{-2}$ m.}
	\label{fig1}
\end{figure*}

We characterized the Marangoni flow of water induced by hydrosoluble surfactants using eight surfactants from the alkyl trimethylammonium halides (C$_n$TABr, $n$=10 to 14; C$_n$TACl, $n$ = 12 and 16) as well as from the sodium alkyl sulfate (C$_n$NaSO$_4$, $n$ = 8 to 12) families (purchased from Sigma-Aldrich before each experimental run, purity 99\%), whose critical micelle concentration varies over two orders of magnitude \cite{Mukerjee1971,Yamanaka1992,Yatcilla1996,Bergeron1997,Hayami1998,Rosen2004,Para2005} (See Supplementary Materials). Surfactant solutions, seeded with light-scattering 10-$\mu$m olive oil droplets, were supplied on the surface of a ultra-pure water layer (Millipore Q, resistivity $\sigma$ = 18.2 M$\Omega$.cm) using a syringe pump (Harvard Apparatus PHD2000) at a constant surfactant molar flow rate $Q_a=\theta Qc$, with $\theta=V_s/(V_s+V_{oil})$ the volume fraction of surfactant solution in the injected liquid, $V_s$ and $V_{oil}$ the volumes of surfactant solution and oil used to prepare the injected solution, $Q$ the total volume flow rate and $c$ the surfactant concentration (Fig. \ref{fig1}a). We made sure that the oil droplets acted as passive tracers and did not influence the properties of the flow (See Supplementary Materials). The flow is divided into three concentric regions (Fig. \ref{fig1}b and movie M3 in Supplementary Materials). A zone of significant light scattering, the source, surrounds the injection point, over distance $r_{s}$. Further downstream, we observed a region of radius $r_{t}$ which exhibits little scattered light that we call the transparent zone. Outside the transparent zone, intense light scattering is observed again, and vortex pairs similar to those reported in the case of thermocapillary Marangoni flows \cite{Mizev2005} grow along the air/water interface and expand outwards. Further from the source, the tracers move only slightly, suggesting that surface tension is spatially homogeneous in this region and that the Marangoni flow is located in the transparent region. A side view of the experiment reveals the existence of a three-dimensional recirculating flow in the bulk fluid below the transparent zone, which changes direction at $r = r_{t}$ and then follows the bottom of the container back towards the source.The slow interfacial vortices might be related to the fate of surfactants at the air/water interface in the outer region \cite{Kovalchuk2006}, which does not have a significant influence on the main flow characteristics relevant to the transparent zone, as we will show later.

\begin{figure}[!ht]
	\includegraphics[scale=1]{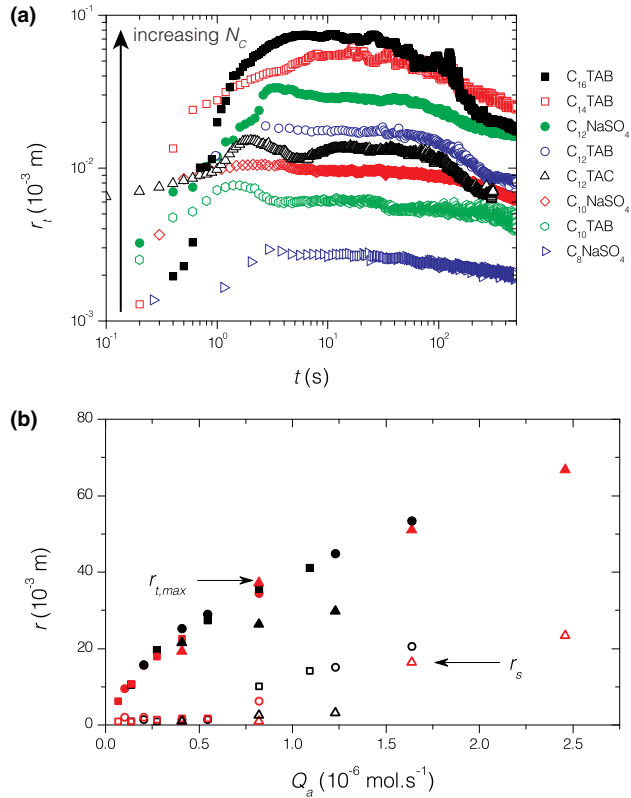}
	\caption{\textbf{Characterization of the transparent zone.} (a) Radius of the transparent zone $r_{t}$ as a function of time $t$ for different surfactants. Surfactant molar flow rate $Q_{a} = 0.52\times10^{-6}$ mol.s$^{-1}$ for all experiments. (b) Maximum radius $r_{t,max}$ (filled symbols) and radius of the source $r_{s}$ (open symbols) as a function of surfactant molar flow rate \textit{Q${}_{a}$}. Data collected for the same amount of injected C$_{12}$NaSO$_4$.}
	\label{fig2}
\end{figure}

The size of the different flow regions depends on the surfactant molecular structure. We observed that, for a constant molar flow rate $Q_{a}$, the radius of the transparent zone $r_{t}$ varies over almost two orders of magnitude when $n$ increases two-fold (Fig. \ref{fig2}a). Also, $r_{t}$ is sensitive to the properties of the surfactant polar headgroup, in particular to its effective radius $r_{eff}$, which takes into account electrostatic and ion-specific effects in its definition \cite{Collins2004,Moreira2010}. For example, an increase of $r_{eff}$ by using C$_{12}$NaSO$_4$ instead of  C$_{12}$TAB, which differ only by their polar headgroups, results in a decrease of $r_t$.

The radius of the transparent zone $r_{t}$ also varies with time (Fig. \ref{fig2}a). After an initial increase at the onset of injection, $r_{t}$ remains constant at a maximal value $r_{t,max}$ for a time dependent on finite-size effects due to the container (see Fig. S2 in Supplementary Materials). Then, $r_{t}$ decreases slowly, before a sharp decrease is observed at longer times, corresponding to a significant increase of the surfactant concentration in the water layer (see Fig. S3 in Supplementary Materials). 

The relationship between the surfactant molar flow rate $Q_{a}$ and $r_{t,max}$ is nonlinear (Fig. \ref{fig2}b) in contrast with the linear dependence between $r_{t,max}$ and \textit{Q${}_{a}$} reported in earlier studies of the continuous Marangoni flow of partially miscible fluids on water \cite{Suciu1967}. The size of the source $r_{s}$ remains equal to the needle diameter until a threshold flow rate is reached, after which $r_s$ increases. The value of the threshold flow rate appears to depend little on the formulation of the injected solution.

\begin{figure}[!ht]
	\includegraphics[scale=1]{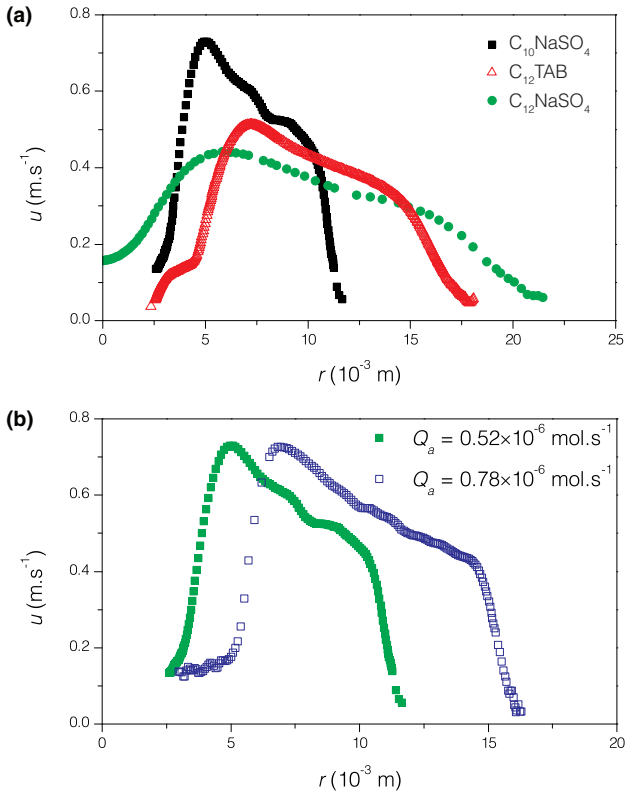}
	\caption{\textbf{Characterization of the velocity field.} (a) Radial component $u$ of the interfacial velocity field in the transparent zone as a function of the radial position $r$.  Surfactant molar flow rate $Q_{a} = 0.52\times10^{-6}$ mol.s$^{-1}$. (b) Radial component $u$ of the interfacial velocity field for a flow induced by C$_{10}$NaSO$_4$ at different molar flow rates $Q_{a}$.}
	\label{fig3}
\end{figure}

To understand the physics underlying the observed flows, we reconstructed the surface velocity field by tracking the interfacial motion of the tracers in the steady regime $r_t=r_{t,max}$ (see movie M4 in Supplementary Materials). The tracers moved along the radial direction only with a velocity $u$ whose shape as a function of $r$ is similar for all the surfactants we tested (Fig. \ref{fig3}a). When the tracers leave the source, where $u\approx 10^{-2}$ m.s$^{-1}$, they accelerate, reach a maximum velocity $u_{max}\approx 0.5$ m.s$^{-1}$, before decelerating as they travel across the transparent area. Finally, tracers decelerate abruptly as they reach $r=r_{t}$. The magnitude of $u_{max}$ decreases with an increase of $n$ and/or $r_{eff}$. The injection flow rate has little effect on the velocity field (Fig. \ref{fig3}b). We note that the shape of the velocity fields is qualitatively similar to those reported in earlier studies on the spreading of partially miscible fluids on water \cite{Suciu1967,Suciu1969,Suciu1970}, though no systematic scaling laws were identified in these earlier works and only partial theoretical descriptions were given.

\begin{figure*}[!ht]
	\includegraphics[scale=1]{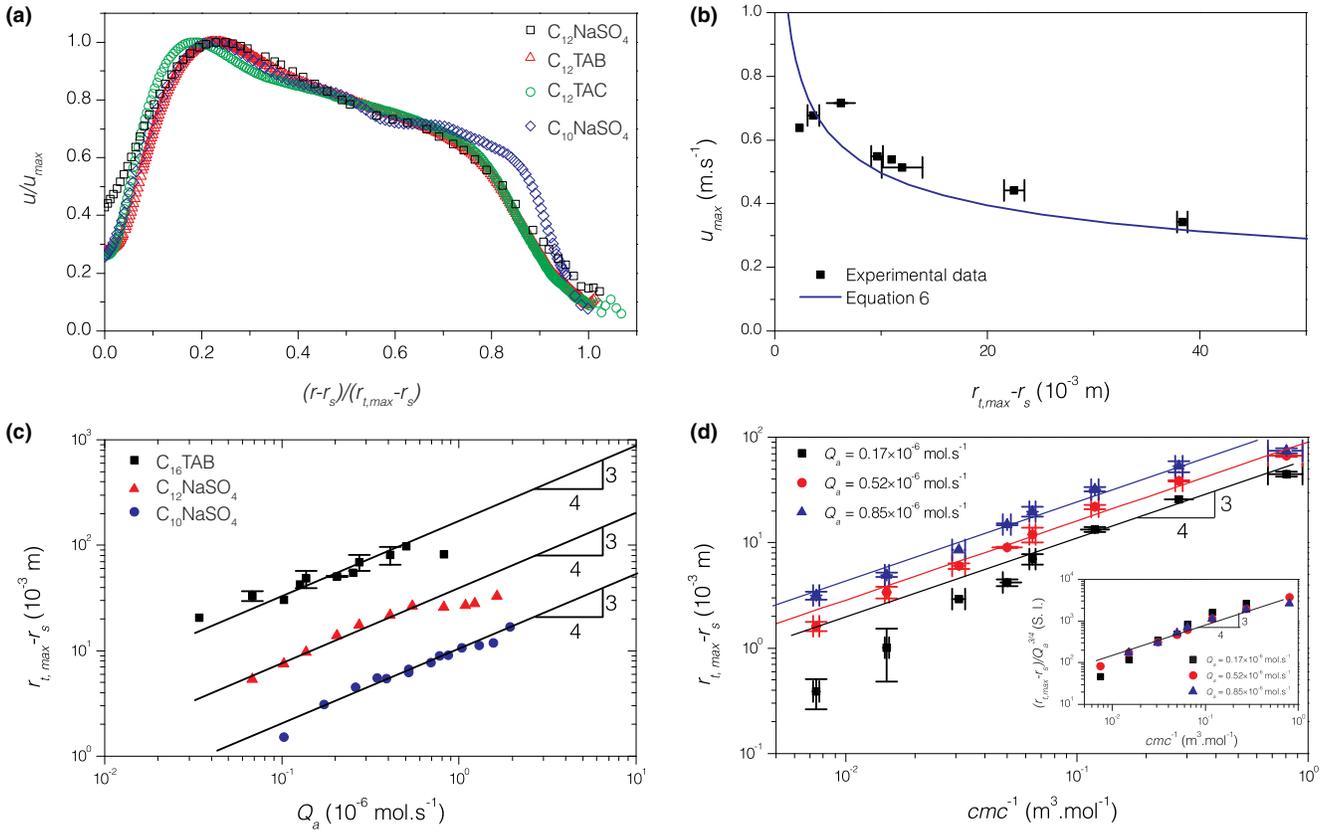}
	\caption{\textbf{Universality of the velocity field in the transparent zone in steady state and scaling laws.} (a) Rescaled velocity profiles $u/u_{max}$ as a function of the rescaled radial coordinate $(r-r_{s})/(r_{t}-r_{s})$ for identical amounts of injected surfactants. (b) Comparison between Eq. \ref{eqn7} and experimental data. (c) Comparison between Eq. \ref{eqn8}  and experimental data for the maximal size of the transparent zone $r_{t,max}-r_{s}$ as a function of the surfactant molar flow rate $Q_{a}$. (d) Comparison between Eq. \ref{eqn8} and experimental data for the maximal size of the transparent zone $r_{t,max}-r_{s}$ as a function of the inverse of the critical micellar concentration $c^*$. Inset: collapse of the experimental data for $r_{t,max}-r_{s}$ when values are rescaled by $Q_{a}^{3/4}$ as a function of $cmc^{-1}$. All data points were measured for the same surfactant amount injected in the layer, $n_{s} = Q_{a}t = 17.2\times10^{-6}$ mol.}
	\label{fig4}
\end{figure*}

Inspired by the similarity of the velocity profiles obtained for surfactants (Fig. 3a), we plotted $u/u_{max}$ versus a rescaled radial coordinate $(r-r_{s})/(r_{t,max}-r_{s})$. Figure \ref{fig4}a shows that the velocity fields obtained for different surfactants collapse on a nearly universal profile when plotted with the rescaled coordinates. The location at which $u = u_{max}$ is $(r-r_{s})/(r_{t,max}-r_{s})\approx 0.2$ for all of the tested surfactants. The profiles have a similar slope during the deceleration stage for $(r-r_{s})/(r_{t,max}-r_{s})>0.2$.

The universality of the velocity fields suggests that a theoretical analysis of the spreading of hydrosoluble surfactants on water in terms of scaling arguments, combining the hydrodynamics of the bulk layer and surfactant physicochemical properties, may capture the features of the flow in the transparent zone. The bulk and the interface of the layer are initially quiescent and surfactant-free. After we begin injecting surfactants, the Marangoni stress induced by the difference between the surface tension of the injected solution and that of ultra-pure water far from the source triggers a flow close to the interface, and momentum diffuses towards the bulk of the layer. In steady state, the balance between convection and diffusion results in a viscous boundary layer with thickness:
 
\begin{eqnarray}
\ell _{\nu} \approx \sqrt{\frac{\nu r^{*} }{u^{*} } },
\label{eqn2}
\end{eqnarray}
with $\nu =\frac{\eta }{\rho }$ the kinematic viscosity, $\eta$ and $\rho$ respectively the dynamic viscosity and the density
of the fluid in the layer, $u^{*}$ a characteristic velocity at the interface and $r^{*}$ the distance over which radial velocity gradients are established, \textit{i.e.} the size of the flow we want to determine. We assume that surface tension gradients in regions extending to $r > r^{*}$ are much smaller than in the area defined by $r<r^{*}$, an assumption supported by previous work on continuous Marangoni flows \cite{Kovalchuk2006}.

The fluid moving along the interface advects surfactants. As there is no surfactant far from the interface, surfactants desorb and diffuse towards the bulk. We assume that adsorption/desorption processes occur on timescales much shorter than the surfactant diffusion in bulk water. Interface-bulk mass exchange is thus diffusion-limited, and a mass transfer boundary layer grows, whose thickness scales as:
\begin{eqnarray}
\ell_{c} \approx \sqrt{\frac{Dr^{*} }{u^{*}}}=Sc^{-1/2} \ell_{\nu},
\label{eqn3}
\end{eqnarray}
with $Sc=\frac{\nu }{D} $ the Schmidt number, which compares the kinematic viscosity $\nu$, \textit{i.e} momentum diffusion constant, to the surfactant bulk diffusion constant ${D}$. Equation \ref{eqn3} is valid if the viscous boundary layer is much larger than the mass transfer boundary layer, \textit{i.e.} if $Sc\gg1$, a condition that is fulfilled in our case as, for a diffusion coefficient $D= 10^{-10}$ m$^{2}$.s$^{-1}$ and $\nu = 10^{-6}$ m$^{2}$.s$^{-1}$ for water, $Sc\approx10^{4}$. The bulk concentration thus varies from a high value just below the interface to zero at the bottom of the mass boundary layer. We choose the cmc of the surfactants as the concentration scale relevant to the description of surfactant transport because of the dependence of the radius of the Marangoni flow on the properties of both the hydrophobic tail and the polar headgroup of the surfactants (Fig. \ref{fig2}a), which are key elements in the thermodynamic definition of the cmc \cite{Moreira2010,Ivanov2007,Tanford1980}.

Our rationale is based on the assumption that the Marangoni flow stops when surfactants injected at the source at a molar flow rate $Q_{a}$ have all desorbed from the interface. Hence, the surfactant mass balance can be expressed as:
\begin{eqnarray} 
Q_{a} \propto r^{*2} D\frac{c^{*}}{\ell_{c} },  
\label{eqn4}
\end{eqnarray} 
with $c^{*}$ the critical micelle concentration. Replacing $\ell_{c}$ by Eq. \ref{eqn3}, we find:
\begin{eqnarray} 
Q_{a} \propto r^{*3/2} (Du^{*} )^{1/2} c^{*}.
\label{eqn5}
\end{eqnarray}
The continuity of stress at the air/water interface writes:
\begin{eqnarray} 
\frac{\eta u^{*} }{\ell_{\nu} } \approx \frac{\gamma _{w} -\gamma _{s} }{r^{*} },  
\label{eqn6}
\end{eqnarray} 
with $\gamma_{w}$ the surface tension of ultra-pure water and $\gamma_{s}$ the surface tension of the surfactant solution. From this condition, we obtain an expression for the velocity:
\begin{eqnarray} 
u^{*} = A\left(\frac{(\gamma_{w} -\gamma_{s} )^{2} }{\eta \rho r^{*} }\right)^{1/3},
\label{eqn7}  
\end{eqnarray}
and by replacing $u^{*}$ in Eq. \ref{eqn5} with Eq. \ref{eqn7}, we obtain:
\begin{eqnarray} 
r^{*} = B\left(\frac{\eta \rho }{(\gamma_{w} -\gamma _{s} )^{2} D^{3} }\right)^{1/8} \left(\frac{Q_{a}}{c^{*}}\right)^{3/4},  
\label{eqn8}
\end{eqnarray} 
where $A$ and $B$ are two dimensionless prefactors. We estimate the values predicted for $u^{*}$ and $r^{*}$ with typical values of the different parameters involved in Eqs. \ref{eqn7} and \ref{eqn8} while assuming that $c^*=10^{-2}$ M, $\gamma_{w} -\gamma _{s}$ is constant for all experiments and equal to 33 mN.m$^{-1}$, a realistic value for the surfactant solutions we used. Setting both $A$ and $B$ to unity, we find $u^{*}\approx$ 0.5 m.s$^{-1}$ and $r^{*}\approx 15\times10^{-3}$ m, which compare very well with our experimental findings for the maximum velocity (Fig. \ref{fig3}).

We compare Eq. \ref{eqn7} to the experimental data by taking $u^{*}= u_{max}$ and $r^{*} = r_{t,max}-r_{s}$. Figure \ref{fig4}b shows that Eq. \ref{eqn7} captures the experimental measurements very well, with a prefactor $A\approx 1$. This agreement supports our assumption of a constant interfacial tension difference $\gamma _{w} -\gamma _{s}$. We note that Eq. \ref{eqn7} fails to capture the data for surfactants forming transparent zones comparable in size to the millimeter-long meniscus connecting the needle tip to water surface, which is not surprising since there is no length scale separation in this case.

The 3/4 exponent of the power law in Eq. \ref{eqn8} is in excellent agreement with the experimental data for $r_{t,max}-r_{s}$ as a function of both $Q_{a}$ and $c^{*}$ (Fig. \ref{fig4}c,d). The prefactor $B$ in Eq. \ref{eqn8} is close to unity (see Fig. S4 in Supplementary Materials). Equation \ref{eqn8} is also able to collapse the experimental data as a function of the cmc onto a master curve (inset in Fig. \ref{fig4}d). The discrepancy between Eq. \ref{eqn8} and data at high flow rates in Fig. \ref{fig4}c is related to the destabilization of the source. Preliminary experiments indicate that the disagreement between data and Eq. \ref{eqn8} at the lowest flow rate in Fig. \ref{fig4}d results from a decrease of the magnitude of $\gamma_w-\gamma_s$. Our experiments confirm that the radius $r_{t}$ of the transparent zone flow increases with an increase of the viscosity of the layer (see Fig. S5 in Supplementary Materials). Thus, the test of the scaling laws against the flow rate $Q_{a}$, the critical micelle concentration of the surfactants and the viscosity of the bulk layer show that Eqs. \ref{eqn7} and \ref{eqn8} contain the appropriate physicochemical ingredients to describe Marangoni flows induced by water-soluble surfactants on water. Moreover, comparison between the scaling laws and the experimental data shows that the values of the prefactors in Eqs. \ref{eqn7} and \ref{eqn8} are close to unity, thus providing further support to the validity of the theoretical arguments. Finally, as all the surfactants we used have similar bulk diffusion coefficients $D$, our results establish the equilibrium cmc as a critical quantity to understand the out-of-equilibrium Marangoni flow.

Our study identifies the link between the molecular structure of hydrosoluble surfactants and the macroscopic Marangoni flow these species induce at an air/water interface. In particular, we demonstrate that this flow has a finite extent whose magnitude is related to the phase behavior of surfactants in water. This connection is important to applications relying on surfactant-induced transport such as emulsification and foaming \cite{Ivanov1997,Cantat2013}, surface coating \cite{Quere1999} and Marangoni drying \cite{Matar2001}. Our findings establish the basis for a new fast method to measure the critical micelle concentration of amphiphiles. Indeed, measuring the cmc often requires the time-consuming measurement of one property of a solution of amphiphiles such as its surface tension as a function of amphiphile concentration. In contrast, our method provides an estimate of the cmc from a single measurement of the size of the spreading area at a given flow rate, accompanied by a single independent measure of the surface tension of the solution.
\begin{acknowledgments}
The authors thank C. Breward, B. Dollet, F. Gilet, M.-C. Jullien, H. Kellay and D. Langevin for helpful discussions. H. A. S. and M. R. thank Unilever Research for partial support of this research.
\end{acknowledgments}


\begin{thebibliography}{99}
\bibitem{Levich1969} V. G. Levich and V. S. Krylov, Annu. Rev. Fluid Mech. \textbf{1}, 293 (1969).
\bibitem{Jobe1993} A. H. Jobe, New Engl. J. Med. \textbf{328}, 861 (1993).
\bibitem{Grotberg1994} J. B. Grotberg, Annu. Rev. Fluid Mech. \textbf{26}, 529 (1994).
\bibitem{Harshey2003} R. M. Harshey, Annu. Rev. Microbiol. \textbf{57}, 249 (2003).
\bibitem{Eisner2005} T. Eisner, M. Eisner and M. Siegler, \textit{Secret Weapons: Defense of Insects, Spiders, Scorpions, and Other Many-Legged Creatures} (Belknap Harvard, Cambridge, Massachusetts, and London, England, 2005).
\bibitem{Bush2006} J. W. M. Bush and D. L. Hu, Annu. Rev. Fluid Mech. \textbf{38}, 339 (2006).
\bibitem{Ivanov1997} I. B. Ivanov and P. A. Kralchevsky, Colloid Surface A \textbf{128}, 155 (1997).
\bibitem{Cantat2013} I. Cantat et al., \textit{Foams: Structure and Dynamics} (Oxford University Press, Oxford, 2013).
\bibitem{Quere1999} D. Qu\'{e}r\'{e}, Annu. Rev. Fluid Mech. \textbf{31}, 347 (1999).
\bibitem{Matar2001} O. K. Matar and R. V. Craster, Phys. Fluids \textbf{13}, 1869 (2001).
\bibitem{Barnes2008} G. T. Barnes, Agr. Water Manage. \textbf{95}, 339 (2008).
\bibitem{Herzig2011} M. A. Herzig, G. T. Barnes and I. R. Gentle, J. Colloid Interf. Sci. \textbf{357}, 239 (2011).
\bibitem{Halpern1992} D. Halpern and J. B. Grotberg, J. Fluid. Mech. \textbf{237}, 1 (1992).
\bibitem{Jensen1993} O. E. Jensen and J. B. Grotberg, Phys. Fluids A \textbf{5}, 58 (1993).
\bibitem{Lee2007} K. S. Lee and V. M. Starov, J. Colloid Interf. Sci. \textbf{314}, 631 (2007).
\bibitem{Lee2009} K. S. Lee, and V. M. Starov, J. Colloid Interf. Sci. \textbf{329}, 361 (2009).
\bibitem{Mukerjee1971} P. Mukerjee, P. and K. J. Mysels, \textit{Critical micelle concentration of aqueous surfactant systems}, No. NSRDS-NBS-36. National Standard Reference Data System, 1971.
\bibitem{Bergeron1997} V. Bergeron, Langmuir \textbf{13}, 3474 (1997).
\bibitem{Hayami1998} Y. Hayami, H. Ichikawa, A. Someya, M. Aratono and K. Motomura, Colloid Polymer Sci. \textbf{276}, 595 (1998).
\bibitem{Para2005} G. Para, E. Jarek snd P. Warszynski, Colloid Surface A \textbf{261}, 65 (2005).
\bibitem{Rosen2004} M. J. Rosen, \textit{Surfactants and Interfacial Phenomena}, 3rd edn (Wiley-Interscience, Hoboken NJ, 2004).
\bibitem{Yamanaka1992} M. Yamanaka, T. Amano, N. Ikeda, M. Aratono, and K. Motomura, Colloid Polymer Sci. \textbf{270}, 682 (1992).
\bibitem{Yatcilla1996} M. T. Yatcilla et al., J. Phys. Chem. \textbf{100}, 5874 (1996).
\bibitem{Mizev2005} A. Mizev, Phys. Fluids \textbf{17}, 122107 (2005).
\bibitem{Kovalchuk2006} N. M. Kovalchuk, and D. Vollhardt, Adv. Colloid Interface Sci. \textbf{120}, 1 (2006).
\bibitem{Collins2004} K. D. Collins, Methods \textbf{34}, 300 (2004).
\bibitem{Moreira2010} L. Moreira and A. Firoozabadi, Langmuir \textbf{26}, 15177 (2010).
\bibitem{Suciu1967} D. G. Suciu, O. Smigelschi, and E. Ruckenstein, AIChE J. \textbf{13}, 1120 (1967).
\bibitem{Suciu1969} D. G. Suciu, O. Smigelschi, and E. Ruckenstein, AIChE J. \textbf{15}, 686 (1969).
\bibitem{Suciu1970} D. G. Suciu, O. Smigelschi, and E. Ruckenstein, J. Colloid Interf. Sci. \textbf{33}, 520 (1970).
\bibitem{Ivanov2007} I. B. Ivanov et al., Adv. Coll. Interfac. Sci. \textbf{134-135}, 105 (2007).
\bibitem{Tanford1980} C. Tanford, \textit{The Hydrophobic Effect: Formation of Micelles and Biological Membranes}, 2nd edn,  (Wiley-Interscience, New York, 1980).
\end{thebibliography}
\end{document}